%% file: paper_LaDoZe.tex
\PassOptionsToPackage{usenames,dvipsnames}{xcolor}
\documentclass[arXiv]{actapoly}

\usepackage[english]{babel}
\usepackage[utf8]{inputenc}
\usepackage{array}
\usepackage{amsfonts}
\usepackage{amsmath}
\usepackage{amssymb}
\usepackage{amsthm}
\usepackage{bm}
\usepackage{soul}
\usepackage[normalem]{ulem}
\usepackage{multirow}
\usepackage{caption}
\usepackage{subcaption}

\makeatletter
\let\r@FirstPage\relax
\let\r@LastPage\relax
\makeatother

\newcolumntype{x}[1]{>{\centering\arraybackslash\hspace{0pt}}p{#1}}

\newcommand{\e}[1]{{\scriptsize\ensuremath{\times 10^{#1}}}}

\mathchardef\mhyphen="2D

\newcommand{\set}[1]{{\mathbb #1}}

\newcommand{\setRn}[1]{\set{R}^{#1}}

\newcommand{\tens}[1]{\bm{#1}}				            

\newcommand{\x}{\tens{x}} 								

\newcommand{\atx}{\ensuremath{(\tens{x})}}

\newcommand{\semtrx}[1]{\mathsf{#1}} 					

\begin{document}

\title[Microstructure reconstruction via ANN]{Microstructure reconstruction via Artificial Neural Networks: A combination of causal and non-causal approach}

\correspondingauthor[K. Latka]{Kryštof Latka}{my}{latka@novyporg.cz}
\author[M. Doškář]{Martin Doškář}{their}
\author[J. Zeman]{Jan Zeman}{their}

\institution{my}{Gymmázium Nový PORG, Pod Krčským lesem 25, Praha 4, Czech Republic}
\institution{their}{Department of Mechanics, Faculty of Civil Engineering, Czech Technical University in Prague, Thákurova 7, 166 29 Praha 6, Czech Republic}

\begin{abstract}

We investigate the applicability of artificial neural networks (ANNs) in reconstructing a sample image of a sponge-like microstructure. We propose to reconstruct the image by predicting the phase of the current pixel based on its causal neighbourhood, and subsequently, use a non-causal ANN model to smooth out the reconstructed image as a form of post-processing. We also consider the impacts of different configurations of the ANN model (e.g., the number of densely connected layers, the number of neurons in each layer, the size of both the causal and non-causal neighbourhood) on the models' predictive abilities quantified by the discrepancy between the spatial statistics of the reference and the reconstructed sample.

\end{abstract}


\maketitle

\section{Introduction}
\label{sect:Intro}

Multi-scale modelling is a powerful predictive tool that bypasses the need for complex constitutive laws by performing auxiliary calculations at lower scales, using a characteristic sample of a material microstructure~\cite{stein_homogenization_2017}.
Such a sample can be easily extracted when analysing a material with regular, periodic arrangement of material's phases; in case of a material with stochastic microstructure, the representative sample is typically constructed artificially such that it matches selected spatial statistics of the material microstructure~\cite{torquato_random_2002}.

While originally the representative samples were generated via optimization approaches, e.g.~\cite{zeman_random_2007}, recently, reconstruction methods relying on machine learning have started to emerge,
using various frameworks including Markov random fields~\cite{Wei_2000}, deep adversarial neural networks~\cite{Yang_2018,Fokina_2020}, or supervised learning using classification trees~\cite{Bostanabad_2016}. Several papers include references to causal and non-causal neighbourhood which both prove to be effective ways of extracting input data for the chosen framework. However, the causal approach generally seems to be the preferred one, with one of these papers even claiming that their model cannot generate valid results if based on a non-causal neighbourhood~\cite{Wei_2000}.

The objective of our work is to reconstruct an image of a microstructure from an almost random noise with microscopic properties as similar as possible to the original image. 
While many of the aforementioned proposed approaches are implementation-complex, we present a simple method using the TensorFlow framework with the Keras sequential API~\cite{TensorFlow}.
We closely follow the methodology of Bostanabad and coworkers, \cite{Bostanabad_2016}; however, instead of using classification trees, we use two distinct Artificial Neural Networks (ANNs), where the first network reconstructs the general pattern of the microstructure and the second network denoises and smoothes out the previously reconstructed pattern. 
For simplicity, we study only two-phase materials, i.e. we test the framework with black and white images.
Our implementation and data are publicly available at a GitLab repository~\cite{gitlab_repo}.

\section{Methodology}
\label{sect:Met}


The proposed reconstruction method comprises two distinct steps:
\begin{enumerate}
    \item The reconstruction of a general shape of the microstructure from a random noise with margins of the reference image used as a seed.
    \item The smoothing procedure of the reconstructed image that improves the local features of the microstructural geometry.
\end{enumerate}

\subsection{Reconstructing microstructural geometry}

\begin{figure}[h]
    \centering
    \scalebox{0.7}{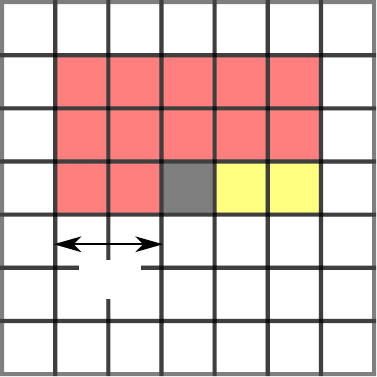}
    \caption{Illustration of the causal neighbourhood with $h^{r} = 2\;\text{pixels}$ around the central pixel highlighted in dark grey. The red pixels represent the input data for the ANN trained in Step 1; however, in order to extract an input structure of a rectangular shape, the black and yellow pixels are also included in the inputs but their values are discarded and replaced by random binary values.}
    \label{fig:causal}
\end{figure}
\begin{figure*}[h!]
\centering
\begin{tabular}{c@{\hspace{4pt}}cc@{\hspace{4pt}}cc@{\hspace{4pt}}c}
	(a) &
		\includegraphics[width=.25\linewidth]{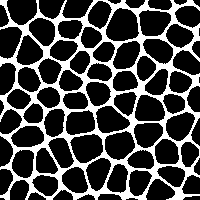} &
	(b) &
		\includegraphics[width=.25\linewidth]{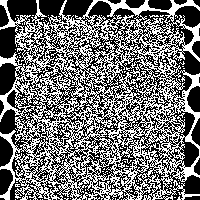} &
	(c) &
		\includegraphics[width=.25\linewidth]{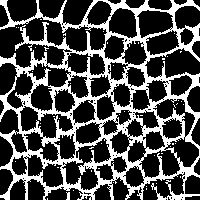}
\end{tabular}
\caption{Microstructure reconstruction process using the causal model. After training the neural network model on input data from the initial microstructure (a), the trained model is used to sequentially predict pixel values in a raster scan order using previously predicted pixel values from the causal neighbourhood.}
\label{fig:reconstruct}
\end{figure*} 
In Step 1, the overall material distribution should be outlined in the reconstructed sample, rendering the key features of the microstructural geometry. 
We opted for a sequential approach, in which new values of individual pixels in the discrete, pixel-like representation of the newly generated microstructural geometry are predicted from the values previously determined at antecedent positions (thus the causal approach), because such an approach has already proven its merits in microstructural reconstruction, cf.~\cite{Bostanabad_2016}.

To this end, we define a rectangular causal neighbourhood of $(2h^{r}+1)\times(h^{r}+1)$ pixels with parameter $h^{r}$ being the given neighbourhood radius; see Fig.~\ref{fig:causal} for an illustration. Note that the positions highlighted in red constitute the real input data; the dark grey and yellow pixels are only a padding (without value) such that the whole input features a regular 2D shape and, consequently, pooling layers can be easily applied. The actual value of the dark grey pixel serves as a label during an extraction of the training data from a reference image.
 
The neural network for predicting the pixel values in Step 1 is designed such that the rectangular input is first subsampled using a $n^{r}_{p} \times n^{r}_{p}$ pooling layer, flattened and passed through several densely connected layers.
Based on our numerical experiments, the maximum pooling layer consistently delivered better results than the average pooling layer. The setup of the training procedure and the effect of the remaining network parameters, namely the number of densely connected layers $n^{r}_{\ell}$, the number of neurons in each layer $n^{r}_{n}$, the size of the pooling layer $n^{r}_{p}$ and the neighbourhood radius $h^{r}$, are discussed later in Section~\ref{sect:Res}. 

Once the network is trained, a new microstructural realization is generated by iterating over pixels of a to-be-reconstructed image in a raster scan order. Consequently, an initial microstructural geometry must be provided in a margin of width $h^{r}$ at the left, top, and bottom edge of the image; the initial values in the remaining part of the image are irrelevant and we generate them as a random binary noise; see Fig.~\ref{fig:reconstruct}b.
 


\subsection{Smoothing procedure}

\begin{figure}[h!]
    \centering
    \scalebox{0.7}{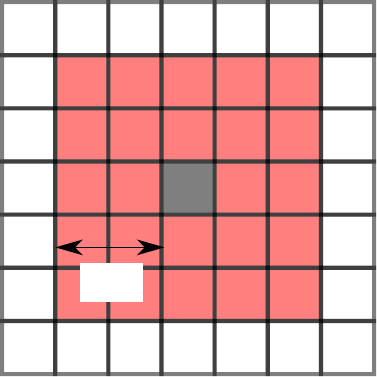}
    \caption{Illustration of the non-causal neighbourhood of a neighbourhood radius of $h^{s} = 2\;\text{pixels}$. The black pixel represents the central pixel and the red pixels represent non-causal neighbourhood. When extracting the neighbourhood, the central black pixel is represented as a random binary value.}
    \label{fig:non-causal}
\end{figure}

Outputs of the model trained in Step 1 usually contain the main features of the trained microstructural geometry; however, local details are typically polluted by random noise; compare Figs.~\ref{fig:reconstruct}a and \ref{fig:reconstruct}c. 
For Step~2 we train an additional neural network to smooth out the image and correct irregularities in the image generated in the Step~1. 

This time, the model works with a complete, i.e. non-causal, square neighbourhood around the central pixel (illustrated in Figure~\ref{fig:non-causal}), usually of a smaller neighbourhood radius $h^{s}$ than in the causal model in Step 1. 
Again, the two-dimensional structure of the input is needed to facilitate a subsampling/pooling layer, which is then followed by flattening and passing through two densely connected layers. The impact of the actual choice (i.e. average vs maximum pooling) of the subsampling layer is discussed in Section~\ref{sect:Res}).

To increase robustness of the trained model and prevent if from learning to simply copy the value of the central pixel, we introduce two errors: (i) the value of the dark grey pixel, which is used as a label in the training, is always randomized in the inputs, and (ii) we also randomly choose value for $\xi\times100\%$ of red pixels from Fig.~\ref{fig:non-causal}.


\begin{figure*}
	\begin{tabular}{c@{\hspace{4pt}}cc@{\hspace{4pt}}c}
		(a) &
        \includegraphics[width=.25\linewidth]{figure02c} &
		(b) &
        \includegraphics[width=.25\linewidth]{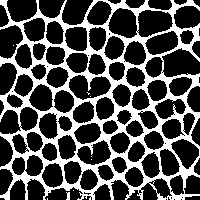} 
	\end{tabular}
    \caption{Effect of the smoothing network based on the non-causal model: (a) reconstructed image serving as an input, (b) microstructure corrected by the smoothing model. Both images are of size 200 x 200 pixels.}
    \label{fig:smooth}
\end{figure*}

\section{Error quantification}
\label{sect:Eq}

In order to assess the performance of the proposed models beyond a visual inspection, we compare generated microstructural samples to the reference images in terms of spatial statistics.
The most straightforward spatial statistics is a volume fraction $\phi$ of a chosen phase (the white phase, i.e. pixels with value 1, in our case).
We define error $\varepsilon^\phi$ as an absolute value of the difference between the volume fraction $\phi^{\text{ref}}$ in the reference sample and the volume fraction $\phi^{\text{gen}}$ in the generated microstructure,

%
%
\begin{equation}
    \varepsilon^\phi=|\phi^\text{ref}-\phi^\text{gen}| \,.
    \label{eq:1}
\end{equation}

The second spatial statistics considered in our work is the two-point probability function $S_{2}\atx$, which states the probability of finding two points separated by $\x$ in a given phase. Since all our data are represented as a regular grid of values, the discrete version of $S_{2}\atx$ can be easily computed using the Fast Fourier Transform~\cite{torquato_random_2002}.
Consequently, we quantify the discrepancy in the reference and a generated microstructure by means of their discrete two-point probability functions $\semtrx{S}_{2}^{\text{ref}}\in\setRn{N_i \times N_j}$ and $\semtrx{S}_{2}^{\text{gen}}\in\setRn{N_i \times N_j}$ as 
%
%
\begin{equation}
    \varepsilon^{S_2}=\frac{\lVert \semtrx{S}_{2}^{\text{gen}} - \semtrx{S}_{2}^{\text{ref}}\rVert_\mathrm{F}}{{N_i N_j}} \,,
    \label{eq:2}
\end{equation}
where $\lVert \semtrx{A} \rVert_\mathrm{F}$ is the Frobenius matrix norm~\cite{golub_matrix_2013}
\begin{equation}
    \lVert \semtrx{A} \rVert^{2}_\mathrm{F}=\sum_{i=1}^{N_i}\sum_{j=1}^{N_j}(A_{i,j})^2 \,.
\end{equation}

Finally, to quantify the effect of our smoothing non-causal model, we add the third error metric $\varepsilon^{D}$ that captures the level of local heterogeneity.
%
%
Assuming a two-phase medium, a microstructure can be represented with a Boolean matrix $\semtrx{M} \in \{0,1\}^{N_i \times N_j}$. For each pixel we can computed a local quantity $D_{i,j}$ as a sum of the averaged absolute differences between the value of the central pixel and its neighbouring eight pixels,
\begin{equation}
    D_{i,j}=\frac{1}{8} \sum_{k,l=-1}^1 |M_{i+k,j+l}-M_{i,j}| \,.
\end{equation}
The error $\varepsilon^{D}$ is then computed again as an average over the image excluding the one-pixel wide margin, i.e.
\begin{equation}
    \varepsilon^{D}=\frac{1}{(N_{i} - 2) (N_{j} - 2)} \sum_{i=2}^{N_{i-1}}\sum_{j=2}^{N_{j-1}} D_{i,j}  \,.
\end{equation}

The reasoning behind this error measure is 
%
that if we consider the phases of pixels in a very small neighbourhood around the central pixel, the number of pixels whose phase is different to that of the central pixel will be lower if the edges are properly smoothed out. Even though this might not necessarily be true for the pixels which form the edge of the reconstructed pattern (and thus, the black and white phase must switch), it will apply on a larger scale (hence, we compute the sum of the values of $D_{ij,}$ for all pixels in the image). Therefore, in theory, the lower the value of $\varepsilon^{D}$ is for an image, the more smoothed out the image should be.

\section{Results}
\label{sect:Res}


We report the effect of parameters on the quality of the microstructural reconstruction quantified with the error measures introduced in the previous chapter.

First, we focus on parameters of the reconstruction model.
\begin{table}[h]
	\centering
	\begin{tabular}{p{0.16\columnwidth}x{0.21\columnwidth}x{0.21\columnwidth}x{0.21\columnwidth}}
		\toprule
		& \multicolumn{3}{c}{Number of neurons $n^{r}_{n}$} \\
		& 9 & 16 & 25 \\\Midrule
		$n^{r}_{\ell} = 2$ & 0.046 & 0.031 & 0.036 \\
		$n^{r}_{\ell} = 3$ & 0.031 & 0.049 & 0.029 \\
		$n^{r}_{\ell} = 4$ & 0.002 & 0.010 & 0.021 \\
		\bottomrule
	\end{tabular}
	\caption{Values of the volume fraction error $\varepsilon^\phi$, depending on the number of layers $n^{r}_{\ell}$ and the number of neurons in each layer $n^{r}_{n}$.}
	\label{tab:recon_volume_1}
\end{table}
\begin{table}[h]
	\centering
	\begin{tabular}{p{0.16\columnwidth}x{0.21\columnwidth}x{0.21\columnwidth}x{0.21\columnwidth}}
			\toprule
		    & \multicolumn{3}{c}{Number of neurons $n^{r}_{n}$} \\
		    & 9 & 16 & 25 \\\Midrule
			$n^{r}_{\ell} = 2$ & 1.36\e{-4} & 9.36\e{-5} & 1.05\e{-4} \\
			$n^{r}_{\ell} = 3$ & 9.32\e{-5} & 1.46\e{-4} & 7.57\e{-5} \\
			$n^{r}_{\ell} = 4$ & 2.96\e{-5} & 3.71\e{-5} & 6.70\e{-5} \\
			\bottomrule
		\end{tabular}
	\caption{Values of the two-point probability error $\varepsilon^{S_2}$, depending on the number of layers $n^{r}_{\ell}$ and the number of neurons in each layer $n^{r}_{n}$.}
	\label{tab:recon_S2_1}
\end{table}
Tables~\ref{tab:recon_volume_1} and~\ref{tab:recon_S2_1} illustrate the impact of altering the number of densely connected layers $n^{r}_{\ell}$ and the number of neurons in each layer $n^{r}_{n}$, while keeping the neighbourhood radius $h^r$ and the pooling size $n^{r}_{p}$ constant, on the volume fraction error $\varepsilon^{\phi}$ and the two-point probability error $\varepsilon^{S_2}$, respectively. In particular, we set $h^r = 15$ and $n^{r}_{p}=3$ as these values produced the most visually appropriate reconstructions in early tests of the model.

The next two tables, Tabs.~\ref{tab:recon_volume_2} and~\ref{tab:recon_S2_2}, summarize the sensitivity study investigating the influence of the neighbourhood radius $h^r$ and the pooling size $n^r_{p}$ on the reconstruction. This time, all the tests were performed with $n^{r}_{\ell} = 2$ densely connected layers with $n^{r}_{n} = 16$ neurons in each layer.
%
\begin{table}[h]
    \centering
    \begin{tabular}{p{0.16\columnwidth}x{0.21\columnwidth}x{0.21\columnwidth}x{0.21\columnwidth}}
		\toprule
		& \multicolumn{3}{c}{Neighbourhood radius $h^{r}$} \\
		& 10 & 12 & 15 \\\Midrule
		$n^{r}_{p} = 2$ & 0.491 & 0.298 & 0.006 \\
		$n^{r}_{p} = 3$ & 0.058 & 0.046 & 0.031 \\
		\bottomrule
	\end{tabular}
    \caption{Values of the volume fraction error $\varepsilon^\phi$, depending on neighbourhood radius $h^{r}$ and pooling size $n^{r}_{p}$.}
    \label{tab:recon_volume_2}
\end{table}
\begin{table}[h]
    \centering
    \begin{tabular}{p{0.16\columnwidth}x{0.21\columnwidth}x{0.21\columnwidth}x{0.21\columnwidth}}
		\toprule
		& \multicolumn{3}{c}{Neighbourhood radius $h^{r}$} \\
		& 10 & 12 & 15 \\\Midrule
		$n^{r}_{p} = 2$ & 2.50\e{-3} & 1.24\e{-3} & 4.05\e{-5} \\
		$n^{r}_{p} = 3$ & 1.81\e{-4} & 1.36\e{-4} & 9.36\e{-5} \\
		\bottomrule
	\end{tabular}
    \caption{Values of the two-point probability error $\varepsilon^{S_2}$, depending on neighbourhood radius $h^{r}$ and pooling size $n^{r}_{p}$.}
    \label{tab:recon_S2_2}
\end{table}


The next three tables, i.e. Tables~\ref{tab:smooth_volume},~\ref{tab:smooth_S2}, and~\ref{tab:smooth_D}, summarize the parametric study for the smoothing model with non-causal neighbourhood.
We chose the reconstructed image obtained by the first model using $n^{r}_{\ell}=2$ layers, each with $n^{r}_{n}=16$ neurons, a neighbourhood radius of $h^{r}=15$ and a pooling size of $n^{r}_{p}=3$ as an input to the model and compared the resulting smoothed-out image to the original microstructure (before reconstruction), recall Fig.~\ref{fig:reconstruct}a, in terms of the error measures introduced in Section~\ref{sect:Res}.
%
%
We carried out two sets of test: one for the average and one for the maximum pooling 2D layer. In each set, we altered the radius of the non-causal neighbourhood $h^{s}$ as well as the magnitude of the artificially introduced noise $\xi$. Each set rendered three tables as we inspected also the level of local heterogeneity $\varepsilon^{D}$, in addition to the errors $\varepsilon^{\phi}$ and $\varepsilon^{S_2}$ already reported for the generative model.
\begin{table}[h]
    \centering
    \begin{tabular}{cp{0.16\columnwidth}x{0.18\columnwidth}x{0.18\columnwidth}x{0.18\columnwidth}}
		\toprule
		& & \multicolumn{3}{c}{Neighbourhood radius $h^{s}$} \\
		& & 5 & 7 & 10 \\\Midrule
		\parbox[t]{2mm}{\multirow{3}{*}{\rotatebox[origin=c]{90}{\small max}}} & $\xi = 0.05$ & 0.041 & 0.002  & 0.040 \\
		& $\xi = 0.10$ & 0.004 & 0.036 & 0.033 \\
		& $\xi = 0.15$ & 0.051 & 0.027 & 0.033 \\
		\Midrule
		\parbox[t]{2mm}{\multirow{3}{*}{\rotatebox[origin=c]{90}{\small average}}} & $\xi = 0.05$ & 0.058 & 0.000  & 0.033 \\
		& $\xi = 0.10$ & 0.037 & 0.003 & 0.030 \\
		& $\xi = 0.15$ & 0.044 & 0.029 & 0.036 \\
		\bottomrule
	\end{tabular}
    \caption{Values of the volume fraction error $\varepsilon^\phi$, depending on the magnitude of artificially introduce noise $\xi$, neighbourhood radius $h^{s}$, and the type of pooling layer used (max or average).}
    \label{tab:smooth_volume}
\end{table}
\begin{table}[h]
    \centering
    \begin{tabular}{cp{0.16\columnwidth}x{0.18\columnwidth}x{0.18\columnwidth}x{0.18\columnwidth}}
		\toprule
		& & \multicolumn{3}{c}{Neighbourhood radius $h^{s}$} \\
		& & 5 & 7 & 10 \\\Midrule
		\parbox[t]{2mm}{\multirow{3}{*}{\rotatebox[origin=c]{90}{\small max}}} & $\xi = 0.05$ & 1.22\e{-4} & 3.15\e{-5} & 1.20\e{-4} \\
		& $\xi = 0.10$ & 3.24\e{-5} & 1.06\e{-4} & 9.88\e{-5} \\
		& $\xi = 0.15$ & 1.23\e{-4} & 7.21\e{-5} & 9.88\e{-5} \\
		\Midrule
		\parbox[t]{2mm}{\multirow{3}{*}{\rotatebox[origin=c]{90}{\small average}}} & $\xi = 0.05$ & 1.76\e{-4} & 3.08\e{-5} & 1.01\e{-4} \\
		& $\xi = 0.10$ & 1.12\e{-4} & 3.25\e{-5} & 8.98\e{-5} \\
		& $\xi = 0.15$ & 1.31\e{-4} & 8.90\e{-5} & 1.08\e{-4} \\
		\bottomrule
	\end{tabular}
    \caption{Values of the two-point probability error $\varepsilon^{S_2}$, depending on the magnitude of artificially introduce noise $\xi$, neighbourhood radius $h^{s}$, and the type of pooling layer used (max or average).}
    \label{tab:smooth_S2}
\end{table}
\begin{table}[h]
    \centering
    \begin{tabular}{cp{0.16\columnwidth}x{0.18\columnwidth}x{0.18\columnwidth}x{0.18\columnwidth}}
		\toprule
		& & \multicolumn{3}{c}{Neighbourhood radius $h^{s}$} \\
		& & 5 & 7 & 10 \\\Midrule
		\parbox[t]{2mm}{\multirow{3}{*}{\rotatebox[origin=c]{90}{\small max}}} & $\xi = 0.05$ & 0.130 & 0.133 & 0.123 \\
		& $\xi = 0.10$ & 0.135 & 0.129 & 0.124 \\
		& $\xi = 0.15$ & 0.123 & 0.130 & 0.127 \\
		\Midrule
		\parbox[t]{2mm}{\multirow{3}{*}{\rotatebox[origin=c]{90}{\small average}}} & $\xi = 0.05$ & 0.113 & 0.120 & 0.112 \\
		& $\xi = 0.10$ & 0.114 & 0.119 & 0.112 \\
		& $\xi = 0.15$ & 0.114 & 0.115 & 0.113 \\
		\bottomrule
	\end{tabular}
    \caption{Values of the local heterogeneity error $\varepsilon^{D}$, depending on the magnitude of artificially introduce noise $\xi$, neighbourhood radius $h^{s}$, and the type of pooling layer used (max or average). For comparison, the value of $\varepsilon^{D}$ for the reference image is 0.097.}
    \label{tab:smooth_D}
\end{table}
\section{Discussion}

First, we add an observation regarding the three error measures defined in Section~\ref{sect:Eq} and our visual perception of the reconstructed microstructures. 
The first error measure, $\varepsilon^\phi$, served as a coarse check that the volume fraction in the reconstructed image is similar to the original; however, it could not assess how similar the reconstructed pattern is to the original. For this purpose, we adopted $\varepsilon^{S_2}$, based on the two-point correlation function, as we excepted it ot be better suited to compare the reconstructed pattern to the original. 
Nevertheless, we noticed a significant discrepancy between the values of $\varepsilon^{S_2}$ for each model and the visual similarity of the reconstructed pattern to the original one. For example, the generative model using $n^{r}_{\ell}=2$ layers, each with $n^{r}_{n}=16$ neurons, neighbourhood radius of $h^{r}=15$ and a pooling size of $n^{r}_{p}=3$, could probably be considered as the most accurate in terms of the visual pattern (the image reconstructed using this model is in Figure~\ref{fig:reconstruct}c), but only seventh best according to $\varepsilon^{S_2}$.
We conclude that other spatial statistics such as two-point cluster function or lineal path should be added to the suite of error measures as well to capture both the global distribution of a microstructure and its local characteristics.
On the other hand, the assessment of the smoothing model by $\varepsilon^{D}$ values was generally in accordance with the quality of the visual appearance of the reconstructed images. 

Our results show that taking more layers with less neurons in each was favourable to the approach with less but more populated layers. Perhaps as expected, the larger neighbourhood radius was considered during training of the generative model, the better the results were. Surprisingly, $n^{r}_{p}=2$ pooling size was better for the largest neighbourhood radius, while larger pooling was preferential in all other cases.

In the case of the smoothing model, considering larger neighbourhood radius $h^{s}$ beyond certain threshold (in our case $h^s = 7$) did not improve its performance. This can be attributed to the different purpose of both models; while the generative model needs information from distant points to properly distribute the material with the sample, the smoothing model is by its nature local. On the other hand, larger pooling layer was consistently outperforming the smaller one in all tests.
Most importantly, we noticed that the values of $\varepsilon^{D}$ were significantly lower for the smoothing models using an average pooling layer instead of a maximum pooling layer. This is probably due to the fact that the average pooling layer ignores sharp features in the image (e. g. individual pixels whose phase was not correctly identified in the Part 1 of the reconstruction), which allowed us to smooth out the edges of the reconstructed pattern in the image.

However, it is important to emphasize that these observations are specific for the considered microstructure.

\section{Conclusions}

Despite the simplicity of the proposed ANN-based model, accompanied by the ease of implementation facilitated by the TensorFlow framework and the Keras Sequential API, the model generates meaningful microstructural geometries.
%
%
The combination of a causal model used for reconstruction and a non-causal smoothing model in particular yielded satisfying results, cf. Figure~\ref{fig:smooth}, considering the fact that the models knew only a limited local information.
We believe that even better results can be obtained by, e.g., incorporating the considered errors directly in the loss function during the training process of individual networks. Yet, this remains to be done in our future work.

The need for the margin of initial values during reconstruction might be seen as a limitation restricting the model to generating microstructural samples only as large as the reference one, from which the margin can be easily copied. 
%
%
However, a possible solution is to take the reference sample, dismember it into pieces and reorder the pieces so that they form the margin of desired size. 
%
Alternatively, starting from different parts of the reference microstructure, a set of smaller samples can be generated; these samples can be then assembled together while blending the microstructure in their overlaps using, e.g., image quilting~\cite{Fokina_2020}.

\begin{acknowledgements}
	M. Doškář and J. Zeman gratefully acknowledge support by the Czech Science Foundation, project No. 19-26143X.
\end{acknowledgements}


\end{document}

%% file: figure01.pdf_tex
\begingroup%
  \makeatletter%
  \providecommand\color[2][]{%
    \errmessage{(Inkscape) Color is used for the text in Inkscape, but the package 'color.sty' is not loaded}%
    \renewcommand\color[2][]{}%
  }%
  \providecommand\transparent[1]{%
    \errmessage{(Inkscape) Transparency is used (non-zero) for the text in Inkscape, but the package 'transparent.sty' is not loaded}%
    \renewcommand\transparent[1]{}%
  }%
  \providecommand\rotatebox[2]{#2}%
  \newcommand*\fsize{\dimexpr\f@size pt\relax}%
  \newcommand*\lineheight[1]{\fontsize{\fsize}{#1\fsize}\selectfont}%
  \ifx\svgwidth\undefined%
    \setlength{\unitlength}{180.56692048bp}%
    \ifx\svgscale\undefined%
      \relax%
    \else%
      \setlength{\unitlength}{\unitlength * \real{\svgscale}}%
    \fi%
  \else%
    \setlength{\unitlength}{\svgwidth}%
  \fi%
  \global\let\svgwidth\undefined%
  \global\let\svgscale\undefined%
  \makeatother%
  \begin{picture}(1,1.00000012)%
    \lineheight{1}%
    \setlength\tabcolsep{0pt}%
    \put(0,0){\includegraphics[width=\unitlength,page=1]{figure01.pdf}}%
    \put(0.28766774,0.22685093){\color[rgb]{0,0,0}\makebox(0,0)[t]{\lineheight{1.25}\smash{\begin{tabular}[t]{c}\Large$h^{r}$\end{tabular}}}}%
  \end{picture}%
\endgroup%

%% file: figure03.pdf_tex
\begingroup%
  \makeatletter%
  \providecommand\color[2][]{%
    \errmessage{(Inkscape) Color is used for the text in Inkscape, but the package 'color.sty' is not loaded}%
    \renewcommand\color[2][]{}%
  }%
  \providecommand\transparent[1]{%
    \errmessage{(Inkscape) Transparency is used (non-zero) for the text in Inkscape, but the package 'transparent.sty' is not loaded}%
    \renewcommand\transparent[1]{}%
  }%
  \providecommand\rotatebox[2]{#2}%
  \newcommand*\fsize{\dimexpr\f@size pt\relax}%
  \newcommand*\lineheight[1]{\fontsize{\fsize}{#1\fsize}\selectfont}%
  \ifx\svgwidth\undefined%
    \setlength{\unitlength}{180.56684479bp}%
    \ifx\svgscale\undefined%
      \relax%
    \else%
      \setlength{\unitlength}{\unitlength * \real{\svgscale}}%
    \fi%
  \else%
    \setlength{\unitlength}{\svgwidth}%
  \fi%
  \global\let\svgwidth\undefined%
  \global\let\svgscale\undefined%
  \makeatother%
  \begin{picture}(1,1.00000048)%
    \lineheight{1}%
    \setlength\tabcolsep{0pt}%
    \put(0,0){\includegraphics[width=\unitlength,page=1]{figure03.pdf}}%
    \put(0.29175605,0.21854346){\color[rgb]{0,0,0}\makebox(0,0)[t]{\lineheight{1.25}\smash{\begin{tabular}[t]{c}\Large$h^{s}$\end{tabular}}}}%
  \end{picture}%
\endgroup%